\documentclass[conference]{IEEEtran}
\IEEEoverridecommandlockouts
\usepackage{cite}
\usepackage{amsmath,amssymb,amsfonts}
\usepackage{algorithmic}
\usepackage{graphicx}
\usepackage{textcomp}
\usepackage{xcolor}
\usepackage{tikz}
\usepackage{booktabs}
\usepackage[bookmarks=false,hidelinks]{hyperref}
\usepackage{pgfplots}
\usepackage{subcaption}
\usetikzlibrary{fit}
\usetikzlibrary{positioning,arrows.meta,backgrounds}
\def\BibTeX{{\rm B\kern-.05em{\sc i\kern-.025em b}\kern-.08em
    T\kern-.1667em\lower.7ex\hbox{E}\kern-.125emX}}
    
\begin{document}

\title{Leveraging Binary Hashing for Scalable Index Assignment in Semantic Communications\\
\thanks{This research was funded in part by the Agence Nationale de la Recherche (ANR) under the ANR-24-IAS1-0002-02 COMSEMA project and the SNS JU 6GARROW project under the EU’s Horizon program Grant Agreement No 101192194, in collaboration with ROK.}
}

\author{
    \IEEEauthorblockN{Leonardo Roque{\textsuperscript{*,$\dagger$}}, Kuangyun Zhu{\textsuperscript{*,$\ddagger$}}, Louis-Adrien Dufrène{\textsuperscript{*}}, \\ Quentin Lampin{\textsuperscript{*}}, Grégoire Lefebvre{\textsuperscript{*}}, Mohamad Assaad{\textsuperscript{$\dagger$}}}
    \IEEEauthorblockA{*Orange Research, France\\
    $\dagger$Laboratoire des Signaux et Systèmes, CentraleSupélec, France\\
    $\ddagger$INSA Lyon, France\\
    Email: \{leonardo.roquealmeidamatos, louisadrien.dufrene, quentin.lampin, gregoire.lefebvre\}@orange.com,\\
    kuangyun.zhu@insa-lyon.fr, mohamad.assaad@centralesupelec.fr}
}

\maketitle

\begin{abstract}
Transmitting semantically structured embedding vectors over digital channels is costly in bit rate, while conventional vector quantization approaches face limitations in scalability, encoding complexity, and index assignment. This paper investigates binary hashing techniques from the information retrieval literature as semantic indexing strategies for general-purpose networks. We study random Locality-Sensitive Hashing (LSH) along with Principal Component Analysis (PCA)-, Canonical Correlation Analysis (CCA)-, and ITQ (Iterative Quantization)-based variants in a semantic communication pipeline over a Binary Symmetric Channel. Embeddings are compressed into binary codes, reconstructed through Hamming retrieval from a shared table, and evaluated on downstream classification. Experiments on the Massive Text Embedding Benchmark (MTEB) and the Banking77 dataset show that data-dependent hashing improves upon random LSH, while ITQ refinement increases robustness to channel noise. Among the evaluated methods, CCA-ITQ achieves the best classification accuracy across the tested channel conditions. These results indicate that binary hashing is an efficient and scalable alternative to quantization-based embedding compression for semantic communications.
\end{abstract}


\section{Introduction}
The increasing number of connected devices and the exponential growth of data traffic have motivated a paradigm shift in communication system design. Classical general-purpose networks focus on the reliable reconstruction of transmitted bits at the receiver, irrespective of their semantic content. In contrast, semantic communications aim to convey only the information relevant to a specific task, thereby reducing the communication overhead by exploiting the meaning of the transmitted data \cite{nguyen2026semanticsurvey}. Recent works have further highlighted the relevance of semantic integration for future practical networks and proposed efficient end-to-end learning architectures that jointly optimize source and channel coding for task-oriented scenarios \cite{xie2021deep}.

As an emerging perspective, semantic communications can be viewed as exchanging information in a form created by and intended for Artificial Intelligence (AI) models rather than as raw source data. Under this AI-native view, embeddings constitute a natural and particularly suitable data format for enabling semantic system design \cite{lampin2024spawc}. Embeddings are high-dimensional, structured vector representations that capture the semantic characteristics of the underlying data. In Natural Language Processing (NLP) in particular, pre-trained sentence embedding models have become a standard tool for representing text in a latent semantic space \cite{muennighoff2023mteb}. However, transmitting raw embeddings is prohibitively expensive in bandwidth-constrained scenarios: a single embedding vector of dimension $d=768$, encoded in IEEE 754 single-precision floating-point format, requires $768 \times 32 = 24{,}576$ bits.
This motivates the need for mapping embeddings to compact binary representations that reduce the bit rate while remaining robust to channel noise, so that semantic similarity is sufficiently preserved for the downstream task.

Compressing embeddings for digital transmission typically involves two steps: quantizing the continuous embedding space into a finite set of discrete vector realizations, and adapting this set of representations for a communication system by assigning a binary index to each discrete realization. One of the fundamental problems in this second step is the Index Assignment Problem (IAP) \cite{wu2011iap}. Given a set of possible discrete vectors, the objective is to assign binary codewords to each of them such that a global distortion metric, for example, the distortion induced by channel errors, is minimized.

Within this quantization-and-indexing framework, existing approaches largely rely on vector quantization (VQ) techniques \cite{gray1984vector,roque2025pq}. These methods first minimize quantization distortion by learning a finite set of representative vectors for the continuous embedding space, and then define a binary index mapping that preserves a form of error locality between the Hamming space and the original embedding space, as in the Binary Switching Algorithm (BSA) \cite{bsa_ref}. While this approach is well studied and locally optimal within the quantization framework, it suffers from three fundamental limitations. First, it introduces an irreducible quantization distortion for any fixed codebook size $K$: reconstruction vectors are artificial centroids that may not correspond to real embeddings, and the reconstruction error cannot be reduced without increasing $K$. Second, assigning the nearest centroid index during transmission requires an encoding complexity that grows as $\mathcal{O}(K d)$, making exhaustive nearest centroid search intractable for large $K$. Third, for performing BSA optimization, the binary mapping refinement incurs a complexity that still grows as $\mathcal{O}(K^2)$. Scaling $K$ to approach the expressiveness of the full training set is therefore computationally prohibitive. There is therefore a need for binary embedding representations that preserve the robustness objectives of index assignment while remaining scalable to large embedding collections.

To address this need, we propose to leverage binary hashing techniques from the information retrieval (IR) literature as a scalable compression framework for semantic communications \cite{wang2018survey}.
Specifically, we investigate Locality-Sensitive Hashing (LSH) \cite{gionis1999similarity}, a family of methods that generates binary codes by thresholding projections of embeddings so that similar vectors remain close in Hamming space (see Section~\ref{sec:hashing:lsh}). We then consider its unsupervised and supervised refinements based on Principal Component Analysis (PCA) \cite{jolliffe2002pca}, Canonical Correlation Analysis (CCA) \cite{hardoon2004canonical}, and Iterative Quantization (ITQ) \cite{gong2011iterative} as binary encoding schemes for embedding transmission over a Binary Symmetric Channel (BSC), where embeddings are encoded directly into binary codes at lower complexity and reconstructed at the receiver by Hamming retrieval from a shared table instead of nearest-centroid quantization.

The main contributions of this paper are threefold: 1) we propose, to the best of our knowledge, the first application of IR binary hashing techniques as dedicated compression schemes for semantic communications, providing a scalable alternative to vector quantization approaches; 2) we benchmark these techniques against uncompressed embeddings and a BSA baseline in a practical semantic communication pipeline over a BSC, across a wide range of bit rates and channel crossover probabilities; 3) we show that PCA- and CCA-based hyperplane selection, refined by ITQ, consistently improves over random LSH, with CCA exploiting label information to better encode the embedding semantic structure that retains classification-relevant information.

\section{System Model}
\label{sec:system}
\subsection{Shared Offline Setup}
The studied system relies on three components that are established offline from a shared labeled training set $\mathcal{D}_{\mathrm{train}} = \{(x_i, y_i)\}_{i=1}^N$, where $N$ is the total number of training samples, and made available at both ends prior to any transmission:
\begin{enumerate}
  \item a fixed pre-trained embedding model $f: \mathcal{X} \to \mathbb{R}^d$,
  \item a binary hasher $H: \mathbb{R}^d \to \{0,1\}^b$ trained on $\{{e}_i = f(x_i)\}_{i=1}^N$,
  \item a lookup database $\mathcal{T}$ (defined below) built by encoding
    the entire training set through~$H$.
\end{enumerate}

No runtime signaling overhead is incurred by this shared knowledge: $\mathcal{T}$ is identical at the transmitter and receiver by construction.

\begin{figure*}[ht]
\centering
\begin{tikzpicture}[
    font=\small,
    >=Stealth,
    arr/.style={->, thick},
    aux/.style={->, thick, dashed},
    block/.style={
        draw,
        rounded corners=2pt,
        minimum width=1.9cm,
        minimum height=0.85cm,
        align=center,
        line width=0.8pt
    },
    db/.style={
        draw,
        rounded corners=2pt,
        minimum width=2.8cm,
        minimum height=0.8cm,
        align=center,
        fill=green!12,
        line width=0.8pt
    }
]

\node (x) at (0,0) {$x$};
\node[block, fill=blue!12, right=0.6cm of x] (enc) {Sentence\\Encoder};
\node[block, fill=cyan!12, right=0.8cm of enc] (hash) {Binary\\Encoder};
\node[block, fill=orange!18, right=0.8cm of hash] (bsc) {BSC$(p)$};
\node[block, fill=violet!12, right=0.8cm of bsc] (dec) {Hamming\\Retrieval};
\node[block, fill=red!10, right=0.8cm of dec] (clf) {Classifier};
\node[right=0.6cm of clf] (yhat) {$\hat y$};

\draw[arr] (x) -- node[above,font=\scriptsize] {} (enc);
\draw[arr] (enc) -- node[above,font=\scriptsize] {$e$} (hash);
\draw[arr] (hash) -- node[above,font=\scriptsize] {$c$} (bsc);
\draw[arr] (bsc) -- node[above,font=\scriptsize] {$\hat c$} (dec);
\draw[arr] (dec) -- node[above,font=\scriptsize] {$\hat e$} (clf);
\draw[arr] (clf) -- (yhat);

\node[db, below=0.5cm of dec] (table) {Shared table $\mathcal{T}$};
\draw[aux] (table.north) -- (dec.south);

\begin{scope}[on background layer]
    \node[draw=blue!50, dashed, rounded corners=4pt, line width=0.8pt,
          fit=(enc)(hash), inner sep=6pt, fill=blue!4,
          label={[font=\small, text=blue!60!black]above:Transmitter}] {};
    \node[draw=red!50, dashed, rounded corners=4pt, line width=0.8pt,
          fit=(dec)(clf)(table), inner sep=6pt, fill=red!3,
          label={[font=\small, text=red!60!black]above:Receiver}] {};
\end{scope}

\end{tikzpicture}
\caption{Studied semantic communication system. The input sentence $x$ is mapped to an embedding $e$, encoded into a binary code $c$, transmitted over a BSC, and reconstructed from a shared table before downstream classification. The Classifier block implements a logistic regression model with a softmax output layer over $77$ banking intent classes (Banking77 dataset).}
\label{fig:system_model}
\end{figure*}
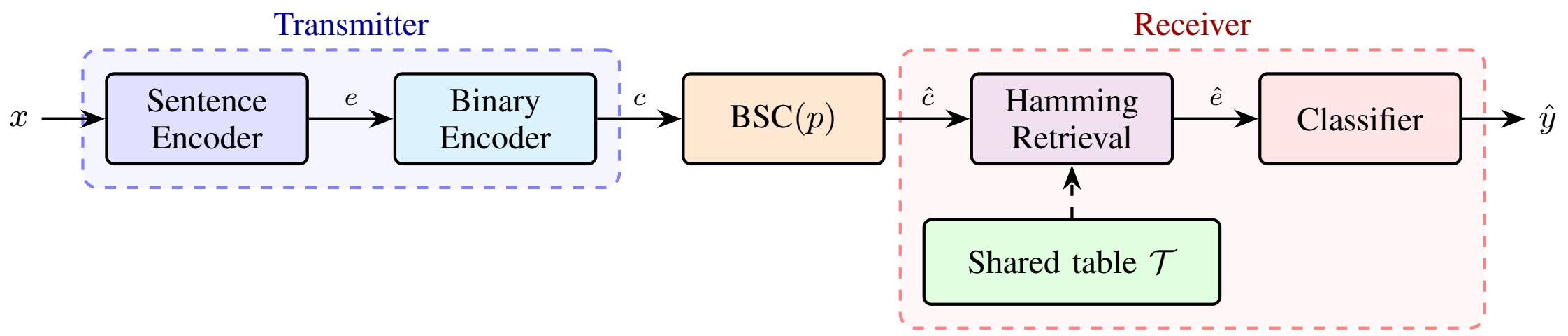

\subsection{Online Communication Pipeline}
The online data flow is illustrated in Fig.~\ref{fig:system_model}. At the transmitter,
a text input $x \in \mathcal{X}$ is mapped to an embedding
${e} = f(x) \in \mathbb{R}^d$ and then compressed to a $b$-bit binary code
\begin{equation}
  {c} = H({e}) \in \{0,1\}^b,
  \label{eq:hash}
\end{equation}
independently of the training set size~$N$.

The code ${c}$ is transmitted over a $\text{BSC}(p)$, which flips each bit
independently with a crossover probability~$p$. At reception, $\hat{c}$ represents the received binary code, potentially accounting for BSC-generated channel distortion.

\subsection{Database and Retrieval Decoder}
\label{sec:ret-dec}
Each training embedding ${e}_i$ is hashed to a binary code ${c}_i = H({e}_i)$. This mapping is not necessarily one-to-one: depending on the number of applied hyperplanes $b$, several embeddings may collide into the same binary code, thereby partitioning the training set into hash cells $\mathcal{S}_{{c}} = \{i : H({e}_i) = {c}\}$. The representative embedding of each populated cell is its mean:
\begin{equation}
  \bar{{e}}_{{c}}
  = \frac{1}{|\mathcal{S}_{{c}}|} \sum_{i \in \mathcal{S}_{{c}}} {e}_i,
  \quad |\mathcal{S}_{{c}}| \geq 1.
  \label{eq:rep}
\end{equation}

Unlike VQ-based schemes where $K$ completely determines the observed quantization error, the hashing-based strategy decreases the intra-cell reconstruction error as~$N$ and~$b$ grow.

Upon receiving~$\hat{{c}}$, the decoder finds the nearest cell codeword by Hamming distance:
\begin{equation}
  {c}^* = \operatorname*{arg\,min}_{{c} \in \mathcal{T}}\;
  d_H(\hat{{c}},\, {c}),
  \label{eq:nn}
\end{equation}
and returns $\hat{{e}} = \bar{{e}}_{{c}^*}$ as the reconstructed embedding. The search in~\eqref{eq:nn} is performed only over the populated cell codewords stored in $\mathcal{T}$, with $|\mathcal{T}| \leq N$, rather than over all $2^b$ possible binary strings. The reconstruction fidelity is quantified by the mean squared error (MSE):
\begin{equation}
  \mathrm{MSE} = \frac{1}{N_{\mathrm{eval}}}
  \sum_{j=1}^{N_{\mathrm{eval}}} \|{e}_j - \hat{{e}}_j\|^2.
  \label{eq:mse}
\end{equation}

\subsection{Downstream Task Evaluation}
Following the MTEB (Massive Text Embedding Benchmark) evaluation protocol~\cite{muennighoff2023mteb}, a logistic regression classifier $\varphi: \mathbb{R}^d \to \mathcal{Y}$ is trained on the uncompressed training embeddings and then applied to the reconstructed embeddings $\hat{{e}}$ at the receiver to isolate the effects of compression and channel distortion on the downstream task. Classification accuracy is the primary metric, evaluated as a service-centric measure and is defined as:
\begin{equation}
    \mathrm{Acc} = \cfrac{N_c}{N_t},
\end{equation}
where $N_c$ represents the number of correct predictions and $N_t$ denotes the total number of predictions made by the classifier. MSE~\eqref{eq:mse} is reported as a secondary reconstruction metric to assess the relationship between embedding fidelity and task performance.

\section{Binary Hashing Schemes}
\label{sec:hashing}

\subsection{General Framework}
\label{sec:hashing:framework}
All schemes share the affine projection structure
\begin{equation}
  H({e}) = \text{sign}\!\bigl({W}({e} - {\mu})\bigr) \in \{-1,+1\}^b,
  \label{eq:proj}
\end{equation}
where $H(e)$ is then mapped to binary bits before transmission, ${W} \in \mathbb{R}^{b \times d}$ is the projection matrix and ${\mu}$ is the training mean vector used to center the observed data as a preprocessing step. Geometrically, each row of ${W}$ defines a hyperplane in the centered embedding space, and the binary code records on which side of each hyperplane the embedding lies. The methods differ in how~${W}$ is obtained: from no data (LSH), from the unlabeled embedding distribution (PCA), or from the joint embedding-label distribution (CCA). The ITQ variants additionally refine the projection with an orthogonal rotation to balance and calibrate bit assignments.

\subsection{Classical LSH}
\label{sec:hashing:lsh}
\begin{figure}[t]
\centering
\begin{tikzpicture}[
    scale=0.75,
    >=Stealth,
    pointq/.style={circle, fill=blue!70!black, inner sep=2pt},
    pointn/.style={circle, fill=green!50!black, inner sep=2pt},
    pointf/.style={circle, fill=red!70!black, inner sep=2pt},
    hyper/.style={thick, gray!75},
    codelabel/.style={font=\scriptsize, fill=white, inner sep=1pt},
    axislabel/.style={font=\scriptsize}
]

\draw[thin] (-2.8,-2.8) rectangle (2.8,2.8);

\draw[->, thin] (-2.8,0) -- (3.0,0) node[right, axislabel] {$z_1$};
\draw[->, thin] (0,-2.8) -- (0,3.0) node[above, axislabel] {$z_2$};

\draw[hyper] (-2.4,-2.4) -- (2.4,2.4);
\draw[hyper] (-2.4,1.2) -- (2.4,-1.2);

\node[font=\scriptsize, gray!70] at (2.10,2.48) {$h_1$};
\node[font=\scriptsize, gray!70] at (2.2,-1.35) {$h_2$};

\node[codelabel] at (-1.65, 1.85) {11}; 
\node[codelabel] at ( 1.65, 1.10) {01}; 
\node[codelabel] at (-1.65,-1.10) {10}; 
\node[codelabel] at ( 1.65,-1.85) {00}; 

\node[pointq, label={[font=\scriptsize, blue!70!black]right:$e$}] at (1.00,0.85) {};
\node[pointn, label={[font=\scriptsize, green!40!black]above:$e'$}] at (1.55,0.35) {};
\node[pointf, label={[font=\scriptsize, red!70!black]left:$e''$}] at (-1.10,-0.85) {};

\draw[dashed, blue!50] (1.00,0.85) -- (1.55,0.35);
\draw[dashed, red!50] (1.00,0.85) -- (-1.10,-0.85);

\end{tikzpicture}
\caption{Illustration of random-projection LSH in a two-dimensional projected space.
The nearby embeddings $e$ and $e'$ fall in the same cell and therefore share the same code, whereas the farther embedding $e''$ lies in a different cell.}
\label{fig:lsh_scheme}
\end{figure}
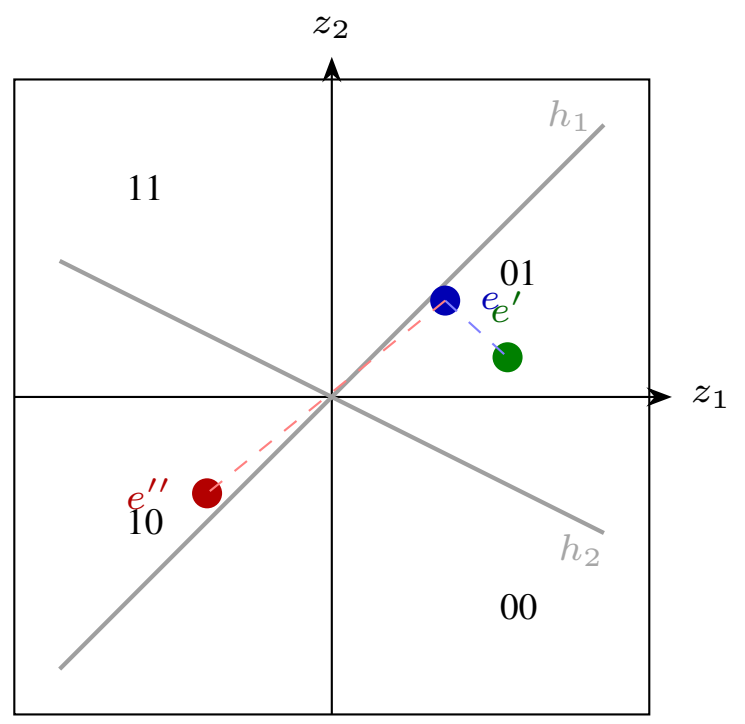

The classical random-projection LSH~\cite{gionis1999similarity} encodes an embedding by testing its position relative to a set of random hyperplanes, as illustrated in Fig.~\ref{fig:lsh_scheme}. Each row of ${W}$ is drawn i.i.d.\ from the Gaussian distribution $\mathcal{N}({0}, {I}_d)$, and the sign of the corresponding projection determines one bit of the code. The final binary representation is obtained by concatenating these $b$ sign bits. Because nearby vectors are more likely to fall on the same side of many hyperplanes, they tend to produce similar binary strings and remain close in Hamming space. Since the projection directions are sampled at random and do not depend on the dataset, random LSH is defined as a data-independent hashing scheme.

\subsection{PCA-Based Hashing}
\label{sec:hashing:pca}
\textbf{PCA-Direct.} Let ${V}_b \in \mathbb{R}^{d \times b}$ contain the top-$b$
eigenvectors of the sample covariance ${\Sigma} = \frac{1}{N}\tilde{{X}}^{\top}\tilde{{X}}$
of the centered training matrix $\tilde{{X}}$. PCA-Direct sets ${W} = {V}_b^{\top}$, projecting along directions of maximum variance over the data distribution. PCA is defined as a task-agnostic, data-dependent unsupervised strategy to perform hyperplane selection in composing the projection matrix $W$.

\textbf{PCA-ITQ.} Iterative Quantization~\cite{gong2011iterative} improves the binarization step by learning an orthogonal rotation ${R} \in \mathbb{R}^{b \times b}$ that minimizes the binary quantization loss
\begin{equation}
  \min_{{R},\, {B}}
  \|{B} - {Z}{R}^{\top}\|_F^2,\quad
  {Z} = \tilde{{X}}{V}_b,
  \label{eq:itq}
\end{equation}
where ${B} \in \{-1,+1\}^{N\times b}$ is the binary code matrix alternately optimized with ${R}$. The role of this rotation is to better align the continuous embedding space with the binary axes before thresholding, so that the sign operation introduces less quantization error than direct binarization. The binary encoder is then given by $H({e}) = \text{sign}({R}{V}_b^{\top}({e}-{\mu}))$.

\subsection{CCA-Based Hashing}
\label{sec:hashing:cca}
CCA~\cite{hardoon2004canonical} constructs a task-oriented, data-dependent projection by maximizing the correlation between embeddings and class labels. The top-$b$ directions are solutions of the regularized generalized eigenvalue problem:
\begin{equation}
  \tilde{{X}}^{\top}\tilde{{Y}}
    \!\left(\tilde{{Y}}^{\top}\tilde{{Y}} + \rho{I}\right)^{-1}\!
    \tilde{{Y}}^{\top}\tilde{{X}}\,{w}_j
  = \lambda_j^2
  \!\left(\tilde{{X}}^{\top}\tilde{{X}} + \rho{I}\right){w}_j,
  \label{eq:cca}
\end{equation}
where $L$ is the number of label classes, $\tilde{{Y}} \in \mathbb{R}^{N \times L}$ is the centered one-hot label matrix and $\rho > 0$ prevents near-singular matrices. The generalized eigenvalues $\lambda_j^2 \in [0,1]$ are the squared canonical correlations; each eigenvector ${w}_j$ is therefore scaled by the corresponding canonical correlation $\lambda_j = \sqrt{\lambda_j^2}$ to form the projection matrix $\hat{{W}} = [\lambda_1{w}_1,\ldots,\lambda_b{w}_b] \in \mathbb{R}^{d \times b}$~\cite{gong2011iterative}.

\textbf{CCA-Direct} defines the projection matrix as ${W} = \hat{{W}}^{\top}$; \textbf{CCA-ITQ}
applies the same ITQ rotation of Section~\ref{sec:hashing:pca} to the
CCA-projected codes ${Z} = \tilde{{X}}\hat{{W}}$.

\section{Experimental Results}
\label{sec:results}

\subsection{Experimental Setup}
\label{sec:results:setup}
\textbf{Dataset and embedding model.}
Experiments are conducted on Banking77\footnote{\url{https://huggingface.co/datasets/mteb/banking77}}, a banking intent classification dataset with $N_{\mathrm{train}}{=}10{,}003$ training samples, $N_{\mathrm{test}}{=}3{,}080$ test samples, and $77$ intent classes. All dataset splits are embedded using the \textit{tomaarsen/mpnet-base-nli}\footnote{\url{https://huggingface.co/tomaarsen/mpnet-base-nli}} model, which produces $d{=}768$-dimensional sentence representations and is used as a general-purpose semantic encoder pre-trained on a larger set of NLP tasks without task-specific fine-tuning~\cite{aarsenembedmodels}.

\textbf{Evaluation protocol.}
Embeddings are standardized using a zero-mean, unit-variance scaler fitted on the training split. Following the MTEB protocol~\cite{muennighoff2023mteb}, each run samples $s{=}8$ training examples per class, yielding $616$ balanced training samples for Banking77. A logistic regression classifier is trained on this subset using uncompressed embeddings and then evaluated at the receiver on reconstructed test embeddings. The procedure is repeated over $n{=}10$ random seeds, and reported accuracies are averaged across runs.

\textbf{Baselines.}
We compare against three baselines: (i)~\emph{Uncompressed} (IEEE~754), which transmits the raw $d{=}768$ \textit{float32} embedding ($24\,576$ bits per embedding); (ii)~\emph{K-means+BSA}, which quantizes embeddings into $K$ centroids and assigns $b{=}\log_2 K$ binary indices with BSA; (iii)~\emph{PQ+BSA}, following~\cite{roque2025pq}, which applies Product Quantization (PQ) with $M$ subspaces and $K$ centroids per subspace, yielding $b{=}M\log_2 K$ bits. Since BSA scales as $\mathcal{O}(K^2)$ per iteration, large K-means codebooks are impractical; in this work, we identify and define $K{=}256$ ($b{=}8$) as the practical ceiling on Banking77. PQ+BSA extends VQ to larger bit budgets with encoding complexity $\mathcal{O}(Kd)$, at the cost of subspace quantization distortion that ignores cross-subspace correlations.

For all BSA-based schemes, we run up to $200$ iterations per round and keep the best mapping among $5$ initializations. The current best assignment is then perturbed by $8$ random permutations, each followed by BSA refinement, and the procedure stops after $4$ consecutive rounds without improvement. BSA distortion evaluation uses a fixed error probability $p{=}0.05$.

\textbf{Bit budgets and channel parameters.}
For accuracy evaluation, hashing schemes are tested with $b \in \{16,32,48,64,80,96,112,128\}$ and $p\in\{0,10^{-3},10^{-2},0.1\}$. Across the same $p$-range, K-means+BSA is evaluated over $b=\{1,2,3,4,5,6,7,8\}$, and PQ+BSA uses $M{=}16$ subspaces with $b=\{1,2,3,4,5,6,7,8\}$ bits per subspace, yielding comparable total bit budgets. Reconstruction MSE is evaluated only for $p{=}0$ to isolate compression-induced distortion.

\subsection{Results and Discussion}
\label{sec:results:discussion}


\pgfplotsset{
  mygrid/.style={
    grid=both,
    grid style={gray!25},
  },
 myaxis/.style={
    width=8cm,
    height=4.8cm,
    xlabel={\small Transmitted bits},
    ylabel={\small Accuracy},
    ymin=0, ymax=1.00,
    xmin=0, xmax=150,
    xtick={0,20,40,60,80,100,120,140},
    ytick={0.0,0.25,0.5,0.75,1.0},
  },
  dashedblack/.style={black,dashed, line width=1pt},
  redtealdash/.style={red, dashed, line width=1pt},
}
\pgfplotsset{compat=1.18}

\begin{figure}[!t]

\pgfplotsset{
  kmstyle/.style={color=brown!85!black, mark=diamond*, mark size=2.3pt, thick},
  kmceilstyle/.style={color=brown!70!black, dotted, line width=1pt},
  bsalimitstyle/.style={color=gray!70, dotted, line width=1pt},
  kmrandstyle/.style={
  color=brown!85!black,
  mark=diamond*,
  mark options={fill=white},
  mark size=2.3pt,
  only marks},
  pqstyle/.style={color=red!75!black, mark=square*, mark size=2.2pt, thick},
  lshstyle/.style={color=black!75, mark=*, mark size=2.0pt, thick},
  pcadirectstyle/.style={color=teal!70!black, mark=triangle*, mark size=2.2pt, thick},
  ccadirectstyle/.style={color=orange!85!black, mark=o, mark size=2.2pt, thick},
  pcaitqstyle/.style={color=blue!70!violet, mark=diamond*, mark size=2.2pt, thick},
  ccaitqstyle/.style={color=magenta!80!black, mark=square*, mark size=2.3pt, thick},
  ieee754style/.style={color=black, dashed, line width=1pt},
  errbars/.style={
    error bars/.cd,
    y dir=both,
    y explicit,
    error bar style={line width=0.85pt, opacity=0.95},
    error mark=none
  }
}

\newcommand{\minmaxcaps}[5]{%
  \draw[color=#1, line width=1.0pt]
    (axis cs:{#2-#5},#3) -- (axis cs:{#2+#5},#3);
  \draw[color=#1, line width=1.0pt]
    (axis cs:{#2-#5},#4) -- (axis cs:{#2+#5},#4);
}

\centering
\begin{tikzpicture}
\begin{axis}[
 title={(a) $p=0$},
 title style={at={(rel axis cs:0.74,0.1)}, anchor=north west},
 myaxis, mygrid, xmax=128, xtick={0,16,32,48,64,80,96,112,128},
 legend columns=4,
 legend style={font=\scriptsize, draw=none, fill=white, fill opacity=0.85,
 text opacity=1, at={(0.5,1.18)}, anchor=south, cells={anchor=west},
 /tikz/every even column/.append style={column sep=0.06cm},
 nodes={inner xsep=0.6pt}}]

\addplot+[kmstyle, errbars] coordinates {
 (1,0.0258)+=(0,0.0002)-=(0,0.0005) (2,0.0447)+=(0,0.0037)-=(0,0.0077)
 (3,0.0919)+=(0,0.0065)-=(0,0.0042) (4,0.1669)+=(0,0.0075)-=(0,0.0078)
 (5,0.3217)+=(0,0.0130)-=(0,0.0094) (6,0.5480)+=(0,0.0065)-=(0,0.0064)
 (7,0.6441)+=(0,0.0121)-=(0,0.0084) (8,0.7197)+=(0,0.0108)-=(0,0.0168)
 (9,0.7675)+=(0,0.0130)-=(0,0.0094) (10,0.7900)+=(0,0.0090)-=(0,0.0114)
 (11,0.7899)+=(0,0.0127)-=(0,0.0194) (12,0.8044)+=(0,0.0089)-=(0,0.0070)};
\addlegendentry{K-means+BSA}

\addplot+[kmrandstyle, forget plot] coordinates {
 (9,0.7675)
 (10,0.7900)
 (11,0.7899)
 (12,0.8044)};
\addlegendimage{thick, color=brown!85!black, mark=diamond*, mark options={fill=white}, mark size=2.3pt}
\addlegendentry{K-means+rand.}

\addplot+[pqstyle, errbars] coordinates {
 (16,0.1328)+=(0,0.0230)-=(0,0.0140) (32,0.2431)+=(0,0.0173)-=(0,0.0269)
 (48,0.3849)+=(0,0.0310)-=(0,0.0213) (64,0.5025)+=(0,0.0209)-=(0,0.0171)
 (80,0.6074)+=(0,0.0173)-=(0,0.0132) (96,0.6895)+=(0,0.0108)-=(0,0.0132)
 (112,0.7438)+=(0,0.0120)-=(0,0.0120) (128,0.7794)+=(0,0.0157)-=(0,0.0171)};
\addlegendentry{PQ+BSA}

\addplot+[kmceilstyle] coordinates {(0,0.8044)(128,0.8044)};
\addlegendentry{K-means ref.}

\addplot+[ieee754style] coordinates {(0,0.7974)(128,0.7974)};
\addlegendentry{IEEE 754}

\addplot+[lshstyle, errbars] coordinates {
 (16,0.2860)+=(0,0.0055)-=(0,0.0049) (32,0.4952)+=(0,0.0100)-=(0,0.0101)
 (48,0.6152)+=(0,0.0062)-=(0,0.0116) (64,0.6581)+=(0,0.0066)-=(0,0.0094)
 (80,0.6875)+=(0,0.0099)-=(0,0.0138) (96,0.7117)+=(0,0.0078)-=(0,0.0074)
 (112,0.7272)+=(0,0.0169)-=(0,0.0126) (128,0.7348)+=(0,0.0080)-=(0,0.0150)};
\addlegendentry{Classical LSH}

\addplot+[pcadirectstyle, errbars] coordinates {
 (16,0.5291)+=(0,0.0106)-=(0,0.0119) (32,0.6768)+=(0,0.0138)-=(0,0.0083)
 (48,0.7219)+=(0,0.0119)-=(0,0.0083) (64,0.7497)+=(0,0.0152)-=(0,0.0143)
 (80,0.7436)+=(0,0.0116)-=(0,0.0180) (96,0.7502)+=(0,0.0134)-=(0,0.0096)
 (112,0.7565)+=(0,0.0088)-=(0,0.0159) (128,0.7565)+=(0,0.0110)-=(0,0.0166)};
\addlegendentry{PCA-Direct}

\addplot+[pcaitqstyle, errbars] coordinates {
 (16,0.5536)+=(0,0.0100)-=(0,0.0069) (32,0.6835)+=(0,0.0090)-=(0,0.0137)
 (48,0.7314)+=(0,0.0098)-=(0,0.0142) (64,0.7521)+=(0,0.0157)-=(0,0.0158)
 (80,0.7691)+=(0,0.0141)-=(0,0.0168) (96,0.7727)+=(0,0.0276)-=(0,0.0194)
 (112,0.7701)+=(0,0.0131)-=(0,0.0071) (128,0.7708)+=(0,0.0061)-=(0,0.0111)};
\addlegendentry{PCA-ITQ}

\addplot+[ccadirectstyle, errbars] coordinates {
 (16,0.5798)+=(0,0.0066)-=(0,0.0058) (32,0.6963)+=(0,0.0095)-=(0,0.0083)
 (48,0.7817)+=(0,0.0170)-=(0,0.0184) (64,0.8011)+=(0,0.0099)-=(0,0.0054)
 (80,0.8071)+=(0,0.0088)-=(0,0.0100) (96,0.7982)+=(0,0.0148)-=(0,0.0183)
 (112,0.7980)+=(0,0.0137)-=(0,0.0100) (128,0.7906)+=(0,0.0069)-=(0,0.0091)};
\addlegendentry{CCA-Direct}

\addplot+[ccaitqstyle, errbars] coordinates {
 (16,0.5699)+=(0,0.0044)-=(0,0.0053) (32,0.7308)+=(0,0.0085)-=(0,0.0172)
 (48,0.7938)+=(0,0.0088)-=(0,0.0077) (64,0.8176)+=(0,0.0084)-=(0,0.0062)
 (80,0.8291)+=(0,0.0056)-=(0,0.0096) (96,0.8271)+=(0,0.0044)-=(0,0.0076)
 (112,0.8288)+=(0,0.0075)-=(0,0.0071) (128,0.8332)+=(0,0.0148)-=(0,0.0073)};
\addlegendentry{CCA-ITQ}

\addplot+[bsalimitstyle] coordinates {(8,0)(8,1)};
\addlegendentry{BSA limit}

\minmaxcaps{brown!85!black}{1}{0.0253}{0.0260}{2.8}
\minmaxcaps{brown!85!black}{2}{0.0370}{0.0484}{2.8}
\minmaxcaps{brown!85!black}{3}{0.0877}{0.0984}{2.8}
\minmaxcaps{brown!85!black}{4}{0.1591}{0.1744}{2.8}
\minmaxcaps{brown!85!black}{5}{0.3123}{0.3347}{2.8}
\minmaxcaps{brown!85!black}{6}{0.5416}{0.5545}{2.8}
\minmaxcaps{brown!85!black}{7}{0.6357}{0.6562}{2.8}
\minmaxcaps{brown!85!black}{8}{0.7029}{0.7305}{2.8}
\minmaxcaps{brown!85!black}{9}{0.7581}{0.7805}{2.8}
\minmaxcaps{brown!85!black}{10}{0.7786}{0.7990}{2.8}
\minmaxcaps{brown!85!black}{11}{0.7705}{0.8026}{2.8}
\minmaxcaps{brown!85!black}{12}{0.7974}{0.8133}{2.8}

\minmaxcaps{red!75!black}{16}{0.1188}{0.1558}{2.8}
\minmaxcaps{red!75!black}{32}{0.2162}{0.2604}{2.8}
\minmaxcaps{red!75!black}{48}{0.3636}{0.4159}{2.8}
\minmaxcaps{red!75!black}{64}{0.4854}{0.5234}{2.8}
\minmaxcaps{red!75!black}{80}{0.5942}{0.6247}{2.8}
\minmaxcaps{red!75!black}{96}{0.6763}{0.7003}{2.8}
\minmaxcaps{red!75!black}{112}{0.7318}{0.7558}{2.8}
\minmaxcaps{red!75!black}{128}{0.7623}{0.7951}{2.8}

\minmaxcaps{black!75}{16}{0.2811}{0.2915}{2.8}
\minmaxcaps{black!75}{32}{0.4851}{0.5052}{2.8}
\minmaxcaps{black!75}{48}{0.6036}{0.6214}{2.8}
\minmaxcaps{black!75}{64}{0.6487}{0.6647}{2.8}
\minmaxcaps{black!75}{80}{0.6737}{0.6974}{2.8}
\minmaxcaps{black!75}{96}{0.7043}{0.7195}{2.8}
\minmaxcaps{black!75}{112}{0.7146}{0.7441}{2.8}
\minmaxcaps{black!75}{128}{0.7198}{0.7428}{2.8}

\minmaxcaps{teal!70!black}{16}{0.5172}{0.5397}{2.8}
\minmaxcaps{teal!70!black}{32}{0.6685}{0.6906}{2.8}
\minmaxcaps{teal!70!black}{48}{0.7136}{0.7338}{2.8}
\minmaxcaps{teal!70!black}{64}{0.7354}{0.7649}{2.8}
\minmaxcaps{teal!70!black}{80}{0.7256}{0.7552}{2.8}
\minmaxcaps{teal!70!black}{96}{0.7406}{0.7636}{2.8}
\minmaxcaps{teal!70!black}{112}{0.7406}{0.7653}{2.8}
\minmaxcaps{teal!70!black}{128}{0.7399}{0.7675}{2.8}

\minmaxcaps{blue!70!violet}{16}{0.5467}{0.5636}{2.8}
\minmaxcaps{blue!70!violet}{32}{0.6698}{0.6925}{2.8}
\minmaxcaps{blue!70!violet}{48}{0.7172}{0.7412}{2.8}
\minmaxcaps{blue!70!violet}{64}{0.7363}{0.7678}{2.8}
\minmaxcaps{blue!70!violet}{80}{0.7523}{0.7832}{2.8}
\minmaxcaps{blue!70!violet}{96}{0.7533}{0.8003}{2.8}
\minmaxcaps{blue!70!violet}{112}{0.7630}{0.7832}{2.8}
\minmaxcaps{blue!70!violet}{128}{0.7597}{0.7769}{2.8}

\minmaxcaps{orange!85!black}{16}{0.5740}{0.5864}{2.8}
\minmaxcaps{orange!85!black}{32}{0.6880}{0.7058}{2.8}
\minmaxcaps{orange!85!black}{48}{0.7633}{0.7987}{2.8}
\minmaxcaps{orange!85!black}{64}{0.7957}{0.8110}{2.8}
\minmaxcaps{orange!85!black}{80}{0.7971}{0.8159}{2.8}
\minmaxcaps{orange!85!black}{96}{0.7799}{0.8130}{2.8}
\minmaxcaps{orange!85!black}{112}{0.7880}{0.8117}{2.8}
\minmaxcaps{orange!85!black}{128}{0.7815}{0.7975}{2.8}

\minmaxcaps{magenta!80!black}{16}{0.5646}{0.5743}{2.8}
\minmaxcaps{magenta!80!black}{32}{0.7136}{0.7393}{2.8}
\minmaxcaps{magenta!80!black}{48}{0.7861}{0.8026}{2.8}
\minmaxcaps{magenta!80!black}{64}{0.8114}{0.8260}{2.8}
\minmaxcaps{magenta!80!black}{80}{0.8195}{0.8347}{2.8}
\minmaxcaps{magenta!80!black}{96}{0.8195}{0.8315}{2.8}
\minmaxcaps{magenta!80!black}{112}{0.8217}{0.8363}{2.8}
\minmaxcaps{magenta!80!black}{128}{0.8259}{0.8480}{2.8}
\end{axis}
\end{tikzpicture}

\centering
\hspace*{-0.66cm}%
\begin{tikzpicture}
\begin{axis}[title={(b) $p=10^{-3}$},
 title style={at={(rel axis cs:0.63,0.1)}, anchor=north west},
 myaxis, mygrid, xmax=128, xtick={0,16,32,48,64,80,96,112,128}]
\addplot+[kmstyle, errbars] coordinates {
 (1,0.0258)+=(0,0.0002)-=(0,0.0005) (2,0.0446)+=(0,0.0038)-=(0,0.0076)
 (3,0.0917)+=(0,0.0064)-=(0,0.0040) (4,0.1662)+=(0,0.0075)-=(0,0.0078)
 (5,0.3189)+=(0,0.0132)-=(0,0.0101) (6,0.5442)+=(0,0.0064)-=(0,0.0065)
 (7,0.6404)+=(0,0.0119)-=(0,0.0079) (8,0.7147)+=(0,0.0103)-=(0,0.0173)};
\addplot+[pqstyle, errbars] coordinates {
 (16,0.1325)+=(0,0.0220)-=(0,0.0143) (32,0.2431)+=(0,0.0166)-=(0,0.0272)
 (48,0.3844)+=(0,0.0305)-=(0,0.0208) (64,0.5014)+=(0,0.0217)-=(0,0.0173)
 (80,0.6064)+=(0,0.0180)-=(0,0.0126) (96,0.6887)+=(0,0.0110)-=(0,0.0134)
 (112,0.7428)+=(0,0.0114)-=(0,0.0120) (128,0.7796)+=(0,0.0162)-=(0,0.0176)};
\addplot+[kmceilstyle] coordinates {(0,0.8044)(128,0.8044)};
\addplot+[ieee754style] coordinates {(0,0.1885)(128,0.1885)};
\addplot+[lshstyle, errbars] coordinates {
 (16,0.2845)+=(0,0.0058)-=(0,0.0062) (32,0.4941)+=(0,0.0095)-=(0,0.0100)
 (48,0.6166)+=(0,0.0058)-=(0,0.0117) (64,0.6579)+=(0,0.0064)-=(0,0.0102)
 (80,0.6873)+=(0,0.0101)-=(0,0.0136) (96,0.7106)+=(0,0.0082)-=(0,0.0074)
 (112,0.7263)+=(0,0.0175)-=(0,0.0117) (128,0.7333)+=(0,0.0076)-=(0,0.0161)};
\addplot+[pcadirectstyle, errbars] coordinates {
 (16,0.5259)+=(0,0.0101)-=(0,0.0117) (32,0.6749)+=(0,0.0140)-=(0,0.0084)
 (48,0.7216)+=(0,0.0122)-=(0,0.0092) (64,0.7505)+=(0,0.0161)-=(0,0.0141)
 (80,0.7435)+=(0,0.0120)-=(0,0.0188) (96,0.7504)+=(0,0.0129)-=(0,0.0098)
 (112,0.7563)+=(0,0.0096)-=(0,0.0154) (128,0.7563)+=(0,0.0113)-=(0,0.0160)};
\addplot+[pcaitqstyle, errbars] coordinates {
 (16,0.5520)+=(0,0.0106)-=(0,0.0072) (32,0.6843)+=(0,0.0089)-=(0,0.0141)
 (48,0.7306)+=(0,0.0096)-=(0,0.0144) (64,0.7516)+=(0,0.0153)-=(0,0.0155)
 (80,0.7689)+=(0,0.0139)-=(0,0.0166) (96,0.7725)+=(0,0.0278)-=(0,0.0186)
 (112,0.7705)+=(0,0.0130)-=(0,0.0071) (128,0.7705)+=(0,0.0058)-=(0,0.0104)};
\addplot+[ccadirectstyle, errbars] coordinates {
 (16,0.5784)+=(0,0.0063)-=(0,0.0063) (32,0.6945)+=(0,0.0097)-=(0,0.0081)
 (48,0.7819)+=(0,0.0178)-=(0,0.0192) (64,0.8015)+=(0,0.0092)-=(0,0.0057)
 (80,0.8067)+=(0,0.0095)-=(0,0.0096) (96,0.7985)+=(0,0.0161)-=(0,0.0190)
 (112,0.7985)+=(0,0.0122)-=(0,0.0099) (128,0.7900)+=(0,0.0080)-=(0,0.0089)};
\addplot+[ccaitqstyle, errbars] coordinates {
 (16,0.5670)+=(0,0.0044)-=(0,0.0057) (32,0.7291)+=(0,0.0083)-=(0,0.0164)
 (48,0.7932)+=(0,0.0094)-=(0,0.0072) (64,0.8178)+=(0,0.0075)-=(0,0.0048)
 (80,0.8293)+=(0,0.0055)-=(0,0.0098) (96,0.8270)+=(0,0.0048)-=(0,0.0076)
 (112,0.8287)+=(0,0.0076)-=(0,0.0070) (128,0.8326)+=(0,0.0141)-=(0,0.0073)};
\addplot+[bsalimitstyle, forget plot] coordinates {(8,0)(8,1)};

\minmaxcaps{brown!85!black}{1}{0.0253}{0.0260}{2.8}
\minmaxcaps{brown!85!black}{2}{0.0370}{0.0484}{2.8}
\minmaxcaps{brown!85!black}{3}{0.0877}{0.0981}{2.8}
\minmaxcaps{brown!85!black}{4}{0.1584}{0.1737}{2.8}
\minmaxcaps{brown!85!black}{5}{0.3088}{0.3321}{2.8}
\minmaxcaps{brown!85!black}{6}{0.5377}{0.5506}{2.8}
\minmaxcaps{brown!85!black}{7}{0.6325}{0.6523}{2.8}
\minmaxcaps{brown!85!black}{8}{0.6974}{0.7250}{2.8}

\minmaxcaps{red!75!black}{16}{0.1182}{0.1545}{2.8}
\minmaxcaps{red!75!black}{32}{0.2159}{0.2597}{2.8}
\minmaxcaps{red!75!black}{48}{0.3636}{0.4149}{2.8}
\minmaxcaps{red!75!black}{64}{0.4841}{0.5231}{2.8}
\minmaxcaps{red!75!black}{80}{0.5938}{0.6244}{2.8}
\minmaxcaps{red!75!black}{96}{0.6753}{0.6997}{2.8}
\minmaxcaps{red!75!black}{112}{0.7308}{0.7542}{2.8}
\minmaxcaps{red!75!black}{128}{0.7620}{0.7958}{2.8}

\minmaxcaps{black!75}{16}{0.2783}{0.2903}{2.8}
\minmaxcaps{black!75}{32}{0.4841}{0.5036}{2.8}
\minmaxcaps{black!75}{48}{0.6049}{0.6224}{2.8}
\minmaxcaps{black!75}{64}{0.6477}{0.6643}{2.8}
\minmaxcaps{black!75}{80}{0.6737}{0.6974}{2.8}
\minmaxcaps{black!75}{96}{0.7032}{0.7188}{2.8}
\minmaxcaps{black!75}{112}{0.7146}{0.7438}{2.8}
\minmaxcaps{black!75}{128}{0.7172}{0.7409}{2.8}

\minmaxcaps{teal!70!black}{16}{0.5142}{0.5360}{2.8}
\minmaxcaps{teal!70!black}{32}{0.6665}{0.6889}{2.8}
\minmaxcaps{teal!70!black}{48}{0.7124}{0.7338}{2.8}
\minmaxcaps{teal!70!black}{64}{0.7364}{0.7666}{2.8}
\minmaxcaps{teal!70!black}{80}{0.7247}{0.7555}{2.8}
\minmaxcaps{teal!70!black}{96}{0.7406}{0.7633}{2.8}
\minmaxcaps{teal!70!black}{112}{0.7409}{0.7659}{2.8}
\minmaxcaps{teal!70!black}{128}{0.7403}{0.7676}{2.8}

\minmaxcaps{blue!70!violet}{16}{0.5448}{0.5626}{2.8}
\minmaxcaps{blue!70!violet}{32}{0.6702}{0.6932}{2.8}
\minmaxcaps{blue!70!violet}{48}{0.7162}{0.7402}{2.8}
\minmaxcaps{blue!70!violet}{64}{0.7361}{0.7669}{2.8}
\minmaxcaps{blue!70!violet}{80}{0.7523}{0.7828}{2.8}
\minmaxcaps{blue!70!violet}{96}{0.7539}{0.8003}{2.8}
\minmaxcaps{blue!70!violet}{112}{0.7634}{0.7835}{2.8}
\minmaxcaps{blue!70!violet}{128}{0.7601}{0.7763}{2.8}

\minmaxcaps{orange!85!black}{16}{0.5721}{0.5847}{2.8}
\minmaxcaps{orange!85!black}{32}{0.6864}{0.7042}{2.8}
\minmaxcaps{orange!85!black}{48}{0.7627}{0.7997}{2.8}
\minmaxcaps{orange!85!black}{64}{0.7958}{0.8107}{2.8}
\minmaxcaps{orange!85!black}{80}{0.7971}{0.8162}{2.8}
\minmaxcaps{orange!85!black}{96}{0.7795}{0.8146}{2.8}
\minmaxcaps{orange!85!black}{112}{0.7886}{0.8107}{2.8}
\minmaxcaps{orange!85!black}{128}{0.7811}{0.7980}{2.8}

\minmaxcaps{magenta!80!black}{16}{0.5613}{0.5714}{2.8}
\minmaxcaps{magenta!80!black}{32}{0.7127}{0.7374}{2.8}
\minmaxcaps{magenta!80!black}{48}{0.7860}{0.8026}{2.8}
\minmaxcaps{magenta!80!black}{64}{0.8130}{0.8253}{2.8}
\minmaxcaps{magenta!80!black}{80}{0.8195}{0.8348}{2.8}
\minmaxcaps{magenta!80!black}{96}{0.8194}{0.8318}{2.8}
\minmaxcaps{magenta!80!black}{112}{0.8217}{0.8363}{2.8}
\minmaxcaps{magenta!80!black}{128}{0.8253}{0.8467}{2.8}
\end{axis}
\end{tikzpicture}

\centering
\hspace*{-0.66cm}%
\begin{tikzpicture}
\begin{axis}[title={(c) $p=10^{-2}$},
 title style={at={(rel axis cs:0.63,0.2)}, anchor=north west},
 myaxis, mygrid, xmax=128, xtick={0,16,32,48,64,80,96,112,128}]
\addplot+[kmstyle, errbars] coordinates {
 (1,0.0249)+=(0,0.0007)-=(0,0.0005) (2,0.0438)+=(0,0.0036)-=(0,0.0078)
 (3,0.0894)+=(0,0.0061)-=(0,0.0050) (4,0.1617)+=(0,0.0081)-=(0,0.0075)
 (5,0.3075)+=(0,0.0113)-=(0,0.0111) (6,0.5149)+=(0,0.0059)-=(0,0.0065)
 (7,0.6006)+=(0,0.0114)-=(0,0.0087) (8,0.6741)+=(0,0.0103)-=(0,0.0166)};
\addplot+[pqstyle, errbars] coordinates {
 (16,0.1307)+=(0,0.0154)-=(0,0.0096) (32,0.2380)+=(0,0.0117)-=(0,0.0221)
 (48,0.3769)+=(0,0.0283)-=(0,0.0227) (64,0.4947)+=(0,0.0199)-=(0,0.0126)
 (80,0.6030)+=(0,0.0155)-=(0,0.0114) (96,0.6838)+=(0,0.0113)-=(0,0.0120)
 (112,0.7409)+=(0,0.0104)-=(0,0.0110) (128,0.7766)+=(0,0.0114)-=(0,0.0191)};
\addplot+[kmceilstyle] coordinates {(0,0.8044)(128,0.8044)};
\addplot+[ieee754style] coordinates {(0,0.0677)(128,0.0677)};
\addplot+[lshstyle, errbars] coordinates {
 (16,0.2698)+=(0,0.0069)-=(0,0.0074) (32,0.4798)+=(0,0.0085)-=(0,0.0087)
 (48,0.6038)+=(0,0.0059)-=(0,0.0110) (64,0.6471)+=(0,0.0052)-=(0,0.0081)
 (80,0.6792)+=(0,0.0092)-=(0,0.0136) (96,0.7012)+=(0,0.0118)-=(0,0.0083)
 (112,0.7256)+=(0,0.0185)-=(0,0.0113) (128,0.7284)+=(0,0.0073)-=(0,0.0138)};
\addplot+[pcadirectstyle, errbars] coordinates {
 (16,0.4944)+=(0,0.0069)-=(0,0.0106) (32,0.6692)+=(0,0.0156)-=(0,0.0097)
 (48,0.7136)+=(0,0.0117)-=(0,0.0091) (64,0.7458)+=(0,0.0155)-=(0,0.0147)
 (80,0.7394)+=(0,0.0129)-=(0,0.0157) (96,0.7456)+=(0,0.0157)-=(0,0.0103)
 (112,0.7529)+=(0,0.0107)-=(0,0.0162) (128,0.7542)+=(0,0.0101)-=(0,0.0152)};
\addplot+[pcaitqstyle, errbars] coordinates {
 (16,0.5224)+=(0,0.0088)-=(0,0.0071) (32,0.6785)+=(0,0.0078)-=(0,0.0149)
 (48,0.7263)+=(0,0.0100)-=(0,0.0143) (64,0.7506)+=(0,0.0153)-=(0,0.0149)
 (80,0.7667)+=(0,0.0132)-=(0,0.0154) (96,0.7693)+=(0,0.0268)-=(0,0.0190)
 (112,0.7697)+=(0,0.0144)-=(0,0.0077) (128,0.7688)+=(0,0.0062)-=(0,0.0113)};
\addplot+[ccadirectstyle, errbars] coordinates {
 (16,0.5528)+=(0,0.0076)-=(0,0.0070) (32,0.6861)+=(0,0.0090)-=(0,0.0075)
 (48,0.7767)+=(0,0.0159)-=(0,0.0182) (64,0.7990)+=(0,0.0098)-=(0,0.0078)
 (80,0.8025)+=(0,0.0105)-=(0,0.0132) (96,0.7928)+=(0,0.0159)-=(0,0.0198)
 (112,0.7940)+=(0,0.0142)-=(0,0.0131) (128,0.7893)+=(0,0.0069)-=(0,0.0097)};
\addplot+[ccaitqstyle, errbars] coordinates {
 (16,0.5390)+=(0,0.0055)-=(0,0.0043) (32,0.7144)+=(0,0.0077)-=(0,0.0163)
 (48,0.7902)+=(0,0.0092)-=(0,0.0064) (64,0.8145)+=(0,0.0063)-=(0,0.0051)
 (80,0.8264)+=(0,0.0064)-=(0,0.0095) (96,0.8288)+=(0,0.0053)-=(0,0.0061)
 (112,0.8275)+=(0,0.0069)-=(0,0.0083) (128,0.8331)+=(0,0.0172)-=(0,0.0068)};
\addplot+[bsalimitstyle, forget plot] coordinates {(8,0)(8,1)};

\minmaxcaps{brown!85!black}{1}{0.0244}{0.0256}{2.8}
\minmaxcaps{brown!85!black}{2}{0.0360}{0.0474}{2.8}
\minmaxcaps{brown!85!black}{3}{0.0844}{0.0955}{2.8}
\minmaxcaps{brown!85!black}{4}{0.1542}{0.1698}{2.8}
\minmaxcaps{brown!85!black}{5}{0.2964}{0.3188}{2.8}
\minmaxcaps{brown!85!black}{6}{0.5084}{0.5208}{2.8}
\minmaxcaps{brown!85!black}{7}{0.5919}{0.6120}{2.8}
\minmaxcaps{brown!85!black}{8}{0.6575}{0.6844}{2.8}

\minmaxcaps{red!75!black}{16}{0.1211}{0.1461}{2.8}
\minmaxcaps{red!75!black}{32}{0.2159}{0.2497}{2.8}
\minmaxcaps{red!75!black}{48}{0.3542}{0.4052}{2.8}
\minmaxcaps{red!75!black}{64}{0.4821}{0.5146}{2.8}
\minmaxcaps{red!75!black}{80}{0.5916}{0.6185}{2.8}
\minmaxcaps{red!75!black}{96}{0.6718}{0.6951}{2.8}
\minmaxcaps{red!75!black}{112}{0.7299}{0.7513}{2.8}
\minmaxcaps{red!75!black}{128}{0.7575}{0.7880}{2.8}

\minmaxcaps{black!75}{16}{0.2624}{0.2767}{2.8}
\minmaxcaps{black!75}{32}{0.4711}{0.4883}{2.8}
\minmaxcaps{black!75}{48}{0.5928}{0.6097}{2.8}
\minmaxcaps{black!75}{64}{0.6390}{0.6523}{2.8}
\minmaxcaps{black!75}{80}{0.6656}{0.6884}{2.8}
\minmaxcaps{black!75}{96}{0.6929}{0.7130}{2.8}
\minmaxcaps{black!75}{112}{0.7143}{0.7441}{2.8}
\minmaxcaps{black!75}{128}{0.7146}{0.7357}{2.8}

\minmaxcaps{teal!70!black}{16}{0.4838}{0.5013}{2.8}
\minmaxcaps{teal!70!black}{32}{0.6595}{0.6848}{2.8}
\minmaxcaps{teal!70!black}{48}{0.7045}{0.7253}{2.8}
\minmaxcaps{teal!70!black}{64}{0.7311}{0.7613}{2.8}
\minmaxcaps{teal!70!black}{80}{0.7237}{0.7523}{2.8}
\minmaxcaps{teal!70!black}{96}{0.7353}{0.7613}{2.8}
\minmaxcaps{teal!70!black}{112}{0.7367}{0.7636}{2.8}
\minmaxcaps{teal!70!black}{128}{0.7390}{0.7643}{2.8}

\minmaxcaps{blue!70!violet}{16}{0.5153}{0.5312}{2.8}
\minmaxcaps{blue!70!violet}{32}{0.6636}{0.6863}{2.8}
\minmaxcaps{blue!70!violet}{48}{0.7120}{0.7363}{2.8}
\minmaxcaps{blue!70!violet}{64}{0.7357}{0.7659}{2.8}
\minmaxcaps{blue!70!violet}{80}{0.7513}{0.7799}{2.8}
\minmaxcaps{blue!70!violet}{96}{0.7503}{0.7961}{2.8}
\minmaxcaps{blue!70!violet}{112}{0.7620}{0.7841}{2.8}
\minmaxcaps{blue!70!violet}{128}{0.7575}{0.7750}{2.8}

\minmaxcaps{orange!85!black}{16}{0.5458}{0.5604}{2.8}
\minmaxcaps{orange!85!black}{32}{0.6786}{0.6951}{2.8}
\minmaxcaps{orange!85!black}{48}{0.7585}{0.7926}{2.8}
\minmaxcaps{orange!85!black}{64}{0.7912}{0.8088}{2.8}
\minmaxcaps{orange!85!black}{80}{0.7893}{0.8130}{2.8}
\minmaxcaps{orange!85!black}{96}{0.7730}{0.8087}{2.8}
\minmaxcaps{orange!85!black}{112}{0.7809}{0.8082}{2.8}
\minmaxcaps{orange!85!black}{128}{0.7796}{0.7962}{2.8}

\minmaxcaps{magenta!80!black}{16}{0.5347}{0.5445}{2.8}
\minmaxcaps{magenta!80!black}{32}{0.6981}{0.7221}{2.8}
\minmaxcaps{magenta!80!black}{48}{0.7838}{0.7994}{2.8}
\minmaxcaps{magenta!80!black}{64}{0.8094}{0.8208}{2.8}
\minmaxcaps{magenta!80!black}{80}{0.8169}{0.8328}{2.8}
\minmaxcaps{magenta!80!black}{96}{0.8227}{0.8341}{2.8}
\minmaxcaps{magenta!80!black}{112}{0.8192}{0.8344}{2.8}
\minmaxcaps{magenta!80!black}{128}{0.8263}{0.8503}{2.8}
\end{axis}
\end{tikzpicture}

\centering
\hspace*{-0.66cm}%
\begin{tikzpicture}
\begin{axis}[title={(d) $p=0.1$},
 title style={at={(rel axis cs:0.63,0.2)}, anchor=north west},
 myaxis, mygrid, xmax=128, xtick={0,16,32,48,64,80,96,112,128}]
\addplot+[kmstyle, errbars] coordinates {
 (1,0.0224)+=(0,0.0026)-=(0,0.0010) (2,0.0363)+=(0,0.0033)-=(0,0.0064)
 (3,0.0664)+=(0,0.0044)-=(0,0.0047) (4,0.1112)+=(0,0.0060)-=(0,0.0057)
 (5,0.1976)+=(0,0.0076)-=(0,0.0099) (6,0.3081)+=(0,0.0039)-=(0,0.0062)
 (7,0.3228)+=(0,0.0051)-=(0,0.0046) (8,0.3554)+=(0,0.0069)-=(0,0.0070)};
\addplot+[pqstyle, errbars] coordinates {
 (16,0.0998)+=(0,0.0064)-=(0,0.0066) (32,0.1719)+=(0,0.0125)-=(0,0.0209)
 (48,0.2956)+=(0,0.0167)-=(0,0.0164) (64,0.3938)+=(0,0.0091)-=(0,0.0136)
 (80,0.5182)+=(0,0.0113)-=(0,0.0078) (96,0.6071)+=(0,0.0124)-=(0,0.0126)
 (112,0.6778)+=(0,0.0089)-=(0,0.0099) (128,0.7075)+=(0,0.0104)-=(0,0.0143)};
\addplot+[kmceilstyle] coordinates {(0,0.8044)(128,0.8044)};
\addplot+[ieee754style] coordinates {(0,0.0784)(128,0.0784)};
\addplot+[lshstyle, errbars] coordinates {
 (16,0.1511)+=(0,0.0035)-=(0,0.0037) (32,0.3118)+=(0,0.0067)-=(0,0.0098)
 (48,0.4406)+=(0,0.0091)-=(0,0.0117) (64,0.5266)+=(0,0.0065)-=(0,0.0078)
 (80,0.5843)+=(0,0.0082)-=(0,0.0083) (96,0.6253)+=(0,0.0105)-=(0,0.0116)
 (112,0.6561)+=(0,0.0156)-=(0,0.0119) (128,0.6782)+=(0,0.0079)-=(0,0.0149)};
\addplot+[pcadirectstyle, errbars] coordinates {
 (16,0.2949)+=(0,0.0044)-=(0,0.0059) (32,0.5054)+=(0,0.0076)-=(0,0.0103)
 (48,0.6195)+=(0,0.0058)-=(0,0.0079) (64,0.6688)+=(0,0.0159)-=(0,0.0162)
 (80,0.6907)+=(0,0.0103)-=(0,0.0193) (96,0.7037)+=(0,0.0106)-=(0,0.0096)
 (112,0.7158)+=(0,0.0069)-=(0,0.0119) (128,0.7236)+=(0,0.0141)-=(0,0.0135)};
\addplot+[pcaitqstyle, errbars] coordinates {
 (16,0.3153)+=(0,0.0074)-=(0,0.0082) (32,0.5537)+=(0,0.0047)-=(0,0.0070)
 (48,0.6520)+=(0,0.0074)-=(0,0.0134) (64,0.7008)+=(0,0.0105)-=(0,0.0145)
 (80,0.7254)+=(0,0.0142)-=(0,0.0137) (96,0.7453)+=(0,0.0203)-=(0,0.0167)
 (112,0.7549)+=(0,0.0116)-=(0,0.0088) (128,0.7545)+=(0,0.0075)-=(0,0.0077)};
\addplot+[ccadirectstyle, errbars] coordinates {
 (16,0.3560)+=(0,0.0044)-=(0,0.0060) (32,0.5644)+=(0,0.0060)-=(0,0.0089)
 (48,0.6997)+=(0,0.0166)-=(0,0.0123) (64,0.7450)+=(0,0.0089)-=(0,0.0112)
 (80,0.7590)+=(0,0.0124)-=(0,0.0119) (96,0.7493)+=(0,0.0127)-=(0,0.0178)
 (112,0.7398)+=(0,0.0102)-=(0,0.0116) (128,0.7433)+=(0,0.0080)-=(0,0.0069)};
\addplot+[ccaitqstyle, errbars] coordinates {
 (16,0.3543)+=(0,0.0058)-=(0,0.0075) (32,0.6162)+=(0,0.0053)-=(0,0.0051)
 (48,0.7443)+=(0,0.0125)-=(0,0.0112) (64,0.7759)+=(0,0.0101)-=(0,0.0081)
 (80,0.8023)+=(0,0.0117)-=(0,0.0091) (96,0.8109)+=(0,0.0121)-=(0,0.0077)
 (112,0.8088)+=(0,0.0078)-=(0,0.0072) (128,0.8230)+=(0,0.0173)-=(0,0.0106)};
\addplot+[bsalimitstyle, forget plot] coordinates {(8,0)(8,1)};

\minmaxcaps{brown!85!black}{1}{0.0214}{0.0250}{2.8}
\minmaxcaps{brown!85!black}{2}{0.0299}{0.0396}{2.8}
\minmaxcaps{brown!85!black}{3}{0.0617}{0.0708}{2.8}
\minmaxcaps{brown!85!black}{4}{0.1055}{0.1172}{2.8}
\minmaxcaps{brown!85!black}{5}{0.1877}{0.2052}{2.8}
\minmaxcaps{brown!85!black}{6}{0.3019}{0.3120}{2.8}
\minmaxcaps{brown!85!black}{7}{0.3182}{0.3279}{2.8}
\minmaxcaps{brown!85!black}{8}{0.3484}{0.3623}{2.8}

\minmaxcaps{red!75!black}{16}{0.0932}{0.1062}{2.8}
\minmaxcaps{red!75!black}{32}{0.1510}{0.1844}{2.8}
\minmaxcaps{red!75!black}{48}{0.2792}{0.3123}{2.8}
\minmaxcaps{red!75!black}{64}{0.3802}{0.4029}{2.8}
\minmaxcaps{red!75!black}{80}{0.5104}{0.5295}{2.8}
\minmaxcaps{red!75!black}{96}{0.5945}{0.6195}{2.8}
\minmaxcaps{red!75!black}{112}{0.6679}{0.6867}{2.8}
\minmaxcaps{red!75!black}{128}{0.6932}{0.7179}{2.8}

\minmaxcaps{black!75}{16}{0.1474}{0.1546}{2.8}
\minmaxcaps{black!75}{32}{0.3020}{0.3185}{2.8}
\minmaxcaps{black!75}{48}{0.4289}{0.4497}{2.8}
\minmaxcaps{black!75}{64}{0.5188}{0.5331}{2.8}
\minmaxcaps{black!75}{80}{0.5760}{0.5925}{2.8}
\minmaxcaps{black!75}{96}{0.6137}{0.6358}{2.8}
\minmaxcaps{black!75}{112}{0.6442}{0.6717}{2.8}
\minmaxcaps{black!75}{128}{0.6633}{0.6861}{2.8}

\minmaxcaps{teal!70!black}{16}{0.2890}{0.2993}{2.8}
\minmaxcaps{teal!70!black}{32}{0.4951}{0.5130}{2.8}
\minmaxcaps{teal!70!black}{48}{0.6116}{0.6253}{2.8}
\minmaxcaps{teal!70!black}{64}{0.6526}{0.6847}{2.8}
\minmaxcaps{teal!70!black}{80}{0.6714}{0.7010}{2.8}
\minmaxcaps{teal!70!black}{96}{0.6941}{0.7143}{2.8}
\minmaxcaps{teal!70!black}{112}{0.7039}{0.7227}{2.8}
\minmaxcaps{teal!70!black}{128}{0.7101}{0.7377}{2.8}

\minmaxcaps{blue!70!violet}{16}{0.3071}{0.3227}{2.8}
\minmaxcaps{blue!70!violet}{32}{0.5467}{0.5584}{2.8}
\minmaxcaps{blue!70!violet}{48}{0.6386}{0.6594}{2.8}
\minmaxcaps{blue!70!violet}{64}{0.6863}{0.7113}{2.8}
\minmaxcaps{blue!70!violet}{80}{0.7117}{0.7396}{2.8}
\minmaxcaps{blue!70!violet}{96}{0.7286}{0.7656}{2.8}
\minmaxcaps{blue!70!violet}{112}{0.7461}{0.7665}{2.8}
\minmaxcaps{blue!70!violet}{128}{0.7468}{0.7620}{2.8}

\minmaxcaps{orange!85!black}{16}{0.3500}{0.3604}{2.8}
\minmaxcaps{orange!85!black}{32}{0.5555}{0.5704}{2.8}
\minmaxcaps{orange!85!black}{48}{0.6874}{0.7163}{2.8}
\minmaxcaps{orange!85!black}{64}{0.7338}{0.7539}{2.8}
\minmaxcaps{orange!85!black}{80}{0.7471}{0.7714}{2.8}
\minmaxcaps{orange!85!black}{96}{0.7315}{0.7620}{2.8}
\minmaxcaps{orange!85!black}{112}{0.7282}{0.7500}{2.8}
\minmaxcaps{orange!85!black}{128}{0.7364}{0.7513}{2.8}

\minmaxcaps{magenta!80!black}{16}{0.3468}{0.3601}{2.8}
\minmaxcaps{magenta!80!black}{32}{0.6111}{0.6215}{2.8}
\minmaxcaps{magenta!80!black}{48}{0.7331}{0.7568}{2.8}
\minmaxcaps{magenta!80!black}{64}{0.7678}{0.7860}{2.8}
\minmaxcaps{magenta!80!black}{80}{0.7932}{0.8140}{2.8}
\minmaxcaps{magenta!80!black}{96}{0.8032}{0.8230}{2.8}
\minmaxcaps{magenta!80!black}{112}{0.8016}{0.8166}{2.8}
\minmaxcaps{magenta!80!black}{128}{0.8124}{0.8403}{2.8}
\end{axis}
\end{tikzpicture}

\caption{Banking77 average classification accuracy versus transmitted bits~$b$,
 for $p\in\{0,\,10^{-3},\,10^{-2},\,0.1\}$ (panels a--d).
 }
\label{fig:acc-eval}
\end{figure}
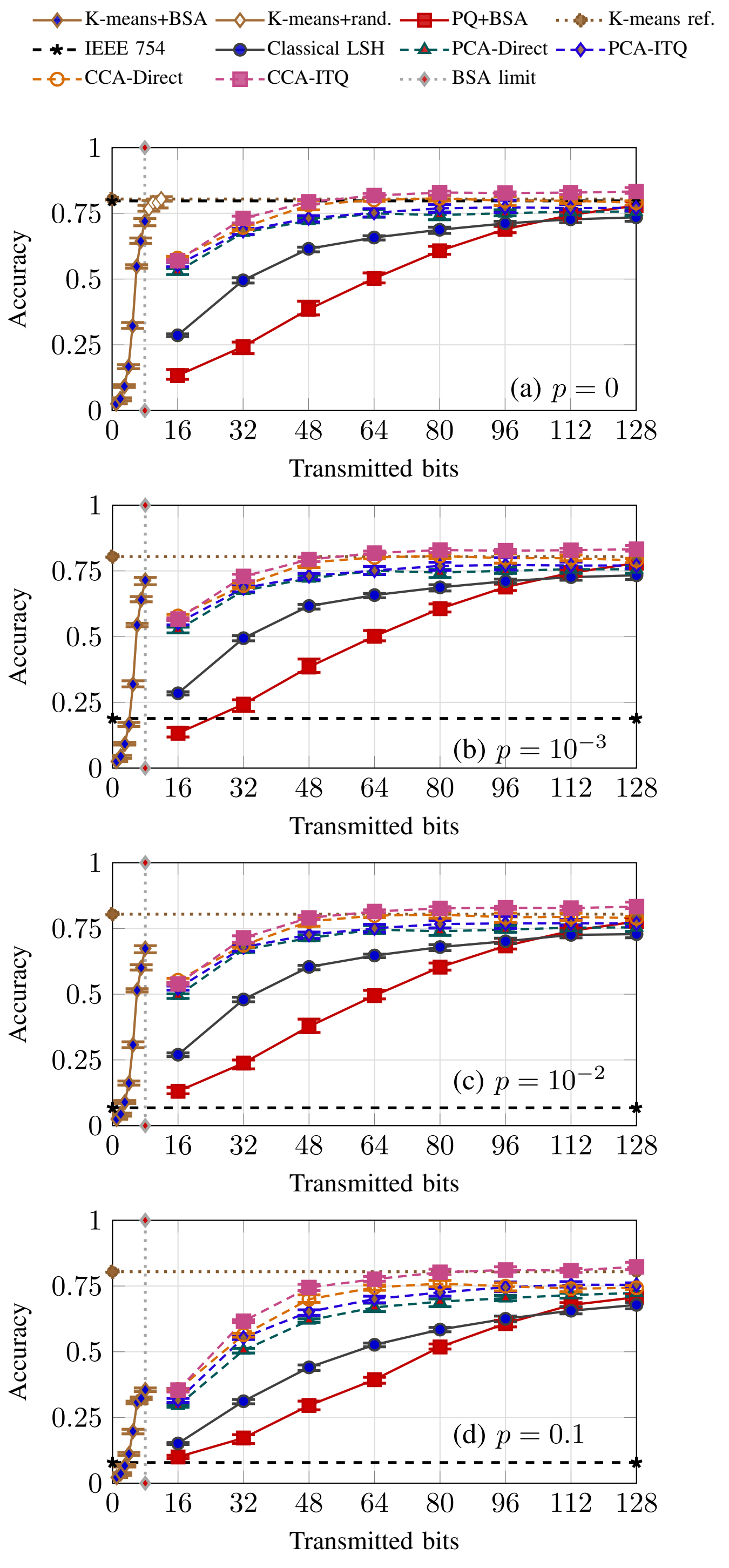

\begin{table}[t]
  \vspace*{0.05in}
  \centering
  \caption{Banking77 10-run average accuracy (\%) and MSE ($\times10^{-4}$, at $p{=}0$).
    K-means+BSA: $K{=}256$ ($b{=}8$, practical ceiling). Others: $b{=}128$.
    \textbf{Bold}: best method per column.}
  \label{tab:results}
  \setlength{\tabcolsep}{3.5pt}
  \begin{tabular}{@{}l c ccc@{}}
    \toprule
    Method & MSE & $p{=}0$ & $p{=}10^{-2}$ & $p{=}0.1$ \\
    \midrule
    IEEE~754 ($b{=}24576$) & \textbf{0} & 79.7 & 6.8 & 7.8 \\
    \midrule
    K-means+BSA ($b{=}8$) & 3.71 & 72.0 & 67.4 & 35.5 \\
    PQ+BSA ($b{=}128$)    & 3.49 & 77.9 & 77.7 & 70.8 \\
    \midrule
    Classical LSH   & 3.52 & 73.5 & 72.8 & 67.8 \\
    PCA-Direct   & 3.28 & 75.7 & 75.4 & 72.4 \\
    PCA-ITQ      & 3.12 & 77.1 & 76.9 & 75.4 \\
    CCA-Direct   & 4.22 & 79.1 & 78.9 & 74.3 \\
    CCA-ITQ      & 5.36 & \textbf{83.3} & \textbf{83.3} & \textbf{82.3} \\
    \bottomrule
  \end{tabular}
\end{table}

Fig.~\ref{fig:acc-eval} reports the 10-run average classification accuracy versus transmitted bits~$b$ on Banking77 for $p\in\{0,10^{-3},10^{-2},0.1\}$ (panels a--d). For K-means+BSA, full BSA optimization is used up to the practical limit $K{=}256$ ($b{=}8$), while larger $K$ values are shown only for $p{=}0$ using random index assignment to isolate the effect of finer quantization; this transition is marked by the dotted vertical line at $b{=}8$. The dotted horizontal line represents as reference the maximum tested K-means accuracy at $p{=}0$, $K{=}4096$, and the dashed horizontal line gives the IEEE~754 uncompressed reference for each $p$. Around each average accuracy over the 10 MTEB runs, the min/max interval bars show that variability remains small overall and indicate stable accuracy trends.

The $p{=}0$ K-means+BSA curve in Fig.~\ref{fig:acc-eval}(a) increases with $K$, from $72.0\%$ at $K{=}256$ to $79.0\%$ at $K{=}1024$ and $80.4\%$ at $K{=}4096$, suggesting that increasing the codebook size effectively reduces reconstruction error and progressively reaches the uncompressed reference result. At the same time, the gain becomes smaller as $K$ grows, which is consistent with the diminishing returns of finer quantization as the representative clusters become less populated and approach the original training distribution. In contrast, the hashing-based schemes already reach competitive or higher accuracies from moderate bit budgets (e.g., in Fig.~\ref{fig:acc-eval}(a) for $b=64$, all hashing-based methods except Classical LSH achieve competitive or better performance than both horizontal references) without requiring nearest-centroid encoding or large codebook optimization. Even though allowing for a much more affordable bit rate scalability to capture quantization expressivity among subspaces, PQ+BSA still observes inferior accuracy performance for moderate bit budgets, and only becomes competitive with hashing schemes at the largest bit budgets (e.g., $b=112,128$).

With regard to robustness against noise, we observe that hashing-based methods better preserve task-relevant structure relative to their error-free performance, even at the highest error probability $p=0.1$ in Fig.~\ref{fig:acc-eval}(d) where remarkably high accuracy results are obtained (e.g., a CCA-ITQ accuracy value at $p=0.1$ similar to the uncompressed reference at $p=0$, using only $128$ bits instead of $24{,}576$ bits, corresponding to a compression factor of $192\times$). Meanwhile, PQ+BSA is slightly more affected by channel distortion and sees a loss of its concurrent zero-error task-relevant structure under these conditions. In contrast, K-means+BSA degrades much more markedly at high channel noise, dropping from $72.0\%$ at $p=0$ to $35.5\%$ at $p=0.1$ at its practical operating point. The uncompressed IEEE~754 baseline is even more sensitive to bit errors, with accuracy collapsing already at very small crossover probabilities because perturbing raw floating-point values severely distorts the received embedding.

Table~\ref{tab:results} reports 10-run average accuracy and reconstruction MSE at the practical operating points of each method: $b{=}128$ for hashing and PQ+BSA, and $b{=}8$ ($K{=}256$) for K-means+BSA. MSE is evaluated only at $p{=}0$ to assess embedding reconstruction capacity, while accuracy is reported for $p\in\{0,10^{-2},0.1\}$. Table~\ref{tab:results} also shows that reconstruction MSE and downstream accuracy are not tightly aligned. CCA-ITQ yields the highest MSE yet the best classification accuracy, whereas PQ+BSA attains lower MSE but lower task performance. Thus, low Euclidean reconstruction error in embedding space does not necessarily preserve the discriminative structure most relevant to the downstream classifier. This is also jointly observed in Fig.~\ref{fig:acc-eval}, where the CCA-based methods provide the strongest overall performance across the tested channel conditions, and notably surpassing the uncompressed IEEE 754 result in Fig.~\ref{fig:acc-eval}(a).

Among the hashing methods, ITQ refinement improves performance consistently over direct projection. At $b{=}128$ and $p{=}0$, the gain is moderate for PCA and more visible for CCA; under noisier conditions, the gap becomes larger. For instance, at $p=0.1$, PCA-ITQ improves over PCA-Direct by 3.0 percentage points, and CCA-ITQ improves over CCA-Direct by 8.0 percentage points. These results are in line with the role of ITQ in producing more balanced binary representations, which appear beneficial not only for binary quantization quality but also for robustness to bit flips.

A similar trend is observed when comparing supervised and unsupervised projections. CCA-based hashing consistently outperforms PCA-based hashing across all channel conditions, highlighting the benefit of using label information to align the binary representation with task-relevant structure. A counterpart is greater task specificity, which may be less suitable in multi-task scenarios. Since the number of informative canonical directions is bounded by $\mathrm{rank}(\tilde{Y})\leq L-1=76$, the CCA curves naturally tend to saturate once $b$ exceeds this effective supervised dimensionality, as observed in Fig.~\ref{fig:acc-eval}.

\section{Complexity and Scalability Discussion}
\label{sec:complexity}
To clarify the practical scalability trade-offs underlying the performance trends observed in Section~\ref{sec:results}, we analyze the computational complexity of the compression schemes across the two main stages of the studied pipeline: offline construction and online operation.

\subsection{Offline Construction Complexity}
For random LSH, building the binary encoder only requires sampling $b$ random projection vectors in $\mathbb{R}^d$, yielding a complexity of $\mathcal{O}(bd)$ \cite{gionis1999similarity}. PCA-based hashing requires estimating principal projection directions from the training embeddings. In a standard dense implementation, the dominant cost comes from covariance formation and eigendecomposition, yielding a complexity on the order of $\mathcal{O}(Nd^2 + d^3)$ \cite{jolliffe2002pca,golub2013matrix}. CCA-based hashing similarly relies on second-order statistics and a generalized eigenvalue problem involving the label space, leading to a comparable but typically heavier offline cost \cite{hardoon2004canonical}. ITQ further adds an alternating orthogonal refinement on the projected $N \times b$ codes, whose complexity scales with the number of refinement iterations \cite{gong2011iterative}.

For the VQ-based baselines, classic K-means training over $K$ centroids costs $\mathcal{O}(I_k N K d)$ over $I_k$ iterations. The subsequent index refinement by BSA or related code-swapping procedures is substantially more demanding, since it explores pairwise code reassignment operations over the codebook and leads to a complexity that grows as $\mathcal{O}(K^2)$ per distortion-evaluation round \cite{bsa_ref}. More generally, optimal index assignment is known to be an NP-hard quadratic assignment problem \cite{wu2011iap}. PQ reduces the dimensionality handled by each sub-quantizer and affords more scalable bit rates, but still requires codebook learning in all $M$ subspaces with index refinement.

\subsection{Online Operation Complexity}
All hashing-based schemes share the same online structure: a $d$-dimensional embedding is projected onto $b$ hyperplanes and thresholded componentwise. Their online encoding complexity is therefore $\mathcal{O}(bd)$, independently of the training set size $N$. This applies to random LSH, PCA-Direct, PCA-ITQ, CCA-Direct, and CCA-ITQ.

By contrast, K-means+BSA requires finding the nearest centroid among $K=2^b$ representatives, which under exhaustive search yields $\mathcal{O}(Kd)$. For PQ+BSA, each of the $M$ sub-vectors is matched against a sub-codebook of size $K$, resulting in an overall complexity of order $\mathcal{O}(MK(d/M))=\mathcal{O}(Kd)$.

For the proposed hashing pipeline, decoding consists of finding the nearest populated binary cell in Hamming space among the entries of the shared table $\mathcal{T}$. Since computing one Hamming distance between two $b$-bit strings costs $\mathcal{O}(b)$, brute-force retrieval over $|\mathcal{T}|$ populated cells requires $\mathcal{O}(|\mathcal{T}|\,b)$, with $|\mathcal{T}| \leq N$. The reconstruction is then a simple table lookup.

For K-means+BSA, decoding is a direct centroid lookup from the received index and therefore costs $\mathcal{O}(1)$. PQ+BSA similarly reconstructs the embedding by concatenating the retrieved sub-centroids from the $M$ shared subcodebooks, yielding then a decoding cost of order $\mathcal{O}(M)$.

\textbf{Discussion.} Regarding the computational trade-off, hashing-based schemes shift complexity toward receiver-side retrieval, while keeping the transmitter complexity low and scalable, with online encoding growing only linearly in the bit budget $b$. In contrast, VQ-based schemes offer very simple decoding once the binary index is received, while incurring heavier offline construction and online encoding costs due to codebook learning, nearest-centroid assignment, and index refinement.

In the Banking77 setting, this difference is reflected by the practical operating points observed in Fig.~\ref{fig:acc-eval}. With $d=768$ and $b=128$, hashing-based encoding scales as $bd = 128 \times 768 \approx 9.8 \times 10^4$, whereas K-means encoding at the practical BSA limit $K=256$ already scales as $Kd = 256 \times 768 \approx 2.0 \times 10^5$, and would rise to about $3.1 \times 10^6$ for $K=4096$ ($b=12$). Hashing therefore provides a substantially lighter transmitter at large bit budgets, at the cost of a receiver-side Hamming search over at most $|\mathcal{T}| \leq N_{\mathrm{train}} = 10{,}003$ populated cells. These orders of magnitude are consistent with the experimental trends: hashing methods remain scalable as $b$ increases and achieve the best overall accuracy--scalability trade-off in our evaluation, especially with CCA-ITQ.

\section{Conclusion}
\label{sec:conclusion}
This paper investigated binary hashing as a semantic compression strategy for transmitting sentence embeddings over a BSC. Experiments on Banking77 showed that data-dependent hashing improves upon random LSH, that ITQ refinement increases robustness under channel noise and that CCA-ITQ achieves the best task results across the tested settings. The complexity analysis further highlighted the favorable scaling of hashing-based encoding compared with VQ-based strategies, despite a more demanding receiver-side retrieval stage.

These findings are limited to a single embedding model and a single downstream task, and broader experiments are needed to assess their generality across a wider range of settings. Future work will therefore consider additional embedding models, larger and more diverse datasets, and other downstream tasks, including multi-task scenarios. Further directions include accelerated Hamming retrieval at the receiver for large or continuously evolving embedding collections and extensions beyond text classification to other modalities, such as image or multimodal embeddings.

\bibliographystyle{IEEEtran}
\bibliography{refs}

\end{document}